\documentclass[12pt]{emulateapj}

\usepackage{color}
\usepackage{graphicx}

\begin{document}


\title{Deep {\it Chandra} Observations of X-ray point sources in M87}

\author{Luan Luan\altaffilmark{1,2,3}}
\author{Christine Jones\altaffilmark{2}}
\author{William R.Forman\altaffilmark{2}}
\author{\'{A}kos Bogd\'{a}n\altaffilmark{2}}
\author{Felipe Andrade-Santos\altaffilmark{2}}
\author{Andy D. Goulding\altaffilmark{4}}
\author{Ryan C. Hickox\altaffilmark{5}}
\author{Meicun Hou\altaffilmark{1,3}}
\author{Zhiyuan Li\altaffilmark{1,3}}
\altaffiliation[1]{ School of Astronomy and Space Science, Nanjing University, Nanjing, 210023, China; luan@smail.nju.edu.cn, lizy@nju.edu.cn}
\altaffiliation[2]{ Harvard-Smithsonian Center for Astrophysics, Cambridge, MA 02138. USA}
\altaffiliation[3]{ Key Laboratory of Morden Astronomy and Astrophysics (Nanjing University), Ministry of Education, Nanjing, 210023, China}
\altaffiliation[4]{ Department Astrophysical Sciences, Princeton University, Princeton, NJ 08544, USA}
\altaffiliation[5]{ Department of Physics and Astronomy, Dartmouth College, Hanover, NH 03755, USA}


\begin{abstract}

\medskip

We present a study of X-ray source populations in M87, the cD galaxy of the Virgo cluster, using 12 archival {\it Chandra} observations with a total exposure of $\sim$680 ks spanning about a decade. A total of 346 point-like sources are detected down to a limiting 0.5--8 keV luminosity of $4\times10^{37}{\rm~erg~s^{-1}}$ and out to a galactocentric radius of $\sim$40 kpc. 
We cross-correlate the X-ray sources with published catalogs of globular clusters (GCs), derived from the ACS Virgo Cluster Survey and the Next Generation Virgo Cluster Survey. 
This results in 122 matches, making it one of the largest samples of GC-hosting X-ray sources in an external galaxy. 
These sources, most likely low-mass X-ray binaries (LMXBs), correspond to $\sim$5\% of all known GCs within the {\it Chandra} field-of-view. 
Conversely, $\sim$50\% of the detected X-ray sources are found in a GC. 
Moreover, red (metal-rich) GCs are $\sim$2.2 times more likely to host an X-ray source than blue (metal-poor) GCs.
We also examine 76 currently known ultra-compact dwarf galaxies around M87, but find no significant X-ray counterparts. 
After statistically accounting for the cosmic X-ray background, we identify $\sim$110 field-LMXBs. 
The GC-LMXBs and field-LMXBs differ in their luminosity function and radial distribution, which indicates that the latter cannot be primarily originated from GCs.
Using another set of deep {\it Chandra} observations toward $\sim$100 kpc northwest of the M87 center, we statistically constrain the abundance of field-LMXBs in the stellar halo, which is consistent with that found in the central region.  
We also identify 40 variable X-ray sources, among which one source is likely a black hole binary residing in a GC.

\smallskip
\end{abstract}


\section{Introduction}
The superb angular resolution and sensitivity afforded by the \textit{Chandra X-ray Observatory} have revolutionized our ability to study discrete X-ray sources in nearby galaxies. 
There is now a consensus that the X-ray sources, presumably low-mass X-ray binaries (LMXBs), account for a substantial, if not dominant, fraction of the total X-ray emission from early-type galaxies (ETGs, including ellipticals and S0s; e.g., \citealp{Sarazin2000}; \citealp{Angelini2001}). 
Many of these X-ray sources are found to reside in globular clusters (GCs), the dense environment of which favors the formation of LMXBs via stellar encounters.  
The fraction of X-ray sources associated with GCs increases with the so-called GC specific frequency, $S_{\rm N} \equiv N_{\rm GC} 10^{0.4({\rm M_V}+15)}$, where $N_{\rm GC}$ is the total number of GCs and ${\rm M_V}$ the absolute V-band magnitude of the host galaxy (\citealp{Harris1991}).   
On the other hand, field-LMXBs (i.e., those located outside GCs) are a good proxy of the host galaxy's stellar mass (\citealp{Gilfanov2004}), with a tendency of higher abundance (i.e., number of sources per unit stellar mass) in younger ETGs (\citealp{Zhang2012}; \citealp{Kim2012}). 
Recent observations provide growing evidence for the presence of ``excess" sources beyond the main distribution of stellar mass, both in relatively isolated, massive ETGs (\citealp{Li2010}; \citealp{Zhang2013};\citealp{van Haaften2018}) and around less massive ETGs in the Virgo cluster (\citealp{Hou2017}).  
The nature of such ``excess" sources is not well understood, but in the case of Virgo, at least some can be attributed to the intra-cluster stellar populations (\citealp{Mihos2017}). 

Residing at the center of Virgo, M87 (=NGC\,4486=VCC\,1316) offers a unique opportunity for studying various stellar populations associated with a cD galaxy. 
In particular, numerous studies have explored the population of GCs in M87, focusing on their spatial and dynamical structure, metallicity distribution and luminosity function (e.g. \citealp{McLaughlin1994}; \citealp{Cohen1998}; \citealp{Harris1998}; \citealp{Kundu1999}; \citealp{Hanes2001}; \citealp{Cote2001}; \citealp{Kissler-Patig2002}; \citealp{Jordan2002}).

The large GC population, together with the large stellar mass, makes M87 an outstanding target for studying the X-ray source populations. 
However, to date few studies have explored this potentially extraordinary X-ray population.  
Based on {\it Chandra} observations with a total exposure of 154\,ks, \citet{Jordan2004} detected X-ray point sources in the inner $\sim$150$\arcsec$ region of M87, and found that (62$\pm$13)\% of these sources can be associated with an optically-identified GC. 
In this work, we employ archival {\it Chandra} observations with a much deeper exposure and larger field-of-view to perform an in-depth study of the X-ray sources in M87, paying special attention to the connection between LMXBs and GCs, as well as the spatial distribution of field-LMXBs.
 
The remainder of this paper is organized as follows. 
In Section 2, we describe the reduction of the {\it Chandra} observations and introduce catalogs of optically-identified GCs in M87. 
In Section 3, we perform X-ray source detection and present the resultant source catalog. 
In Section 4, we analyze the X-ray source properties and their connection with GCs.
In Section 5 we summarize and discuss our results.  
Throughout this work, we adopt a distance of 16 Mpc for M87 (\citealp{Tonry2001}; 1" corresponds to a linear scale of 78 pc).
We quote errors at the 1\,$\sigma$ (68.3\%) confidence level, unless otherwise stated.  

\begin{figure}
\epsscale{1.0}
\includegraphics[width=0.5\textwidth]{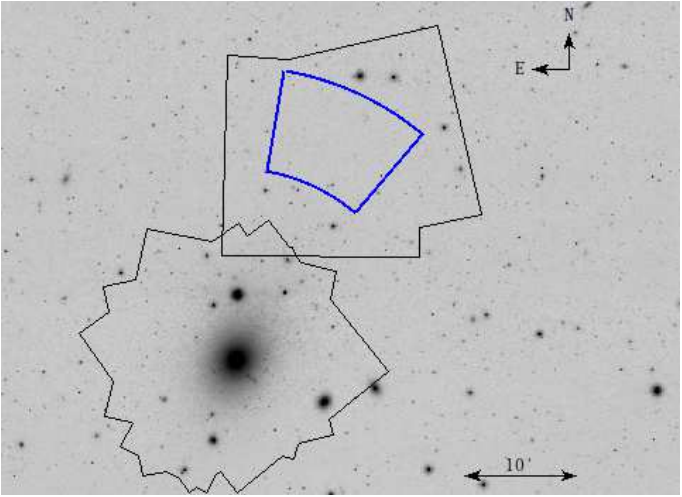}
\caption{A Sloan Digital Sky Survey g-band image showing M87 and its vicinity.  The field-of-view of {\it Chandra} observations covering the main stellar content of M87 (i.e., the Center-Field) is outlined by the lower left polygon, while the Off-Field {\it Chandra} observations are outlined by the upper right polygon. The blue sector illustrates the region where source surface density is examined (Section 4.3).}
\label{fig:fov}
\end{figure}

\begin{figure*}
\epsscale{1.0}
\includegraphics[width=7in]{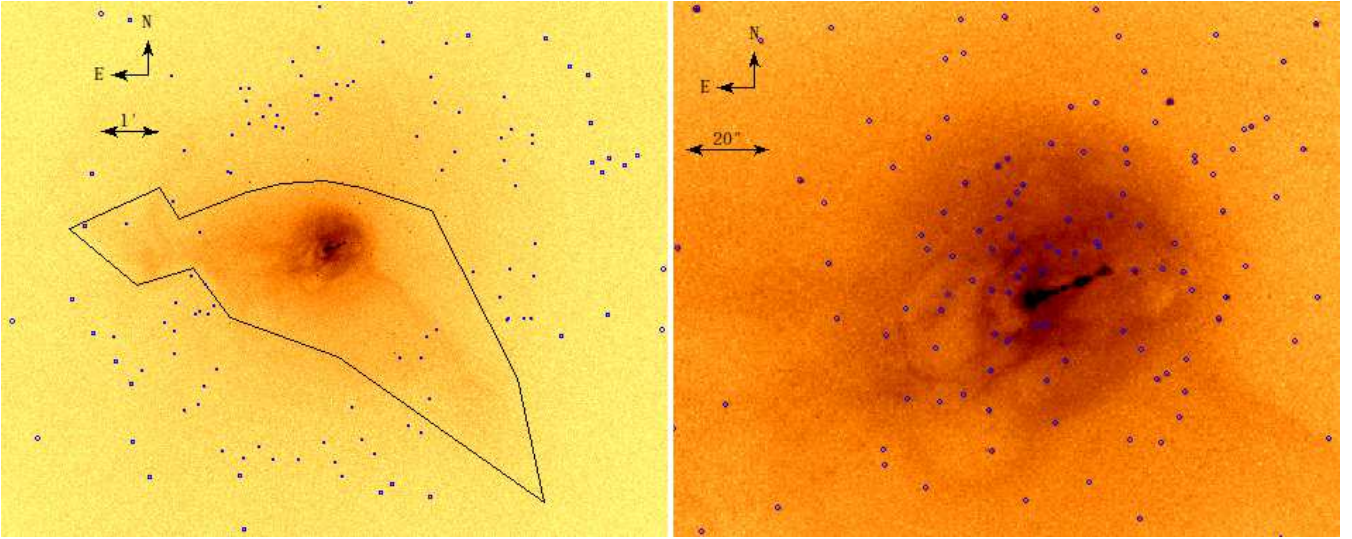}

\caption{Both panels show the 0.5-8 keV flux images of M87. 
The detected X-ray point sources are marked by circles with a 90\% enclosed-energy radius. The {\it left panel} shows a field-of-view of $10\arcmin\times10\arcmin$, while the {\it right panel} is a zoom-in view of the M87 core region. The polygon in the {\it left panel} outlines regions where the M87 jet and diffuse ``clumps" of X-ray emission produce spurious sources, which are removed from the final source catalog.}
\label{fig:fluximage}
\end{figure*}

\section{Data Preparation}
\subsection{{\it Chandra} observations}
The central region of M87 has been frequently observed by {\it Chandra}.
In this work, we utilized 12 {\it Chandra} observations taken with the Advanced CCD Imaging Spectrometer (ACIS), which all have an exposure longer than 10 ks and are publicly available as of 2016. Among them, four were taken with ACIS-S and the remaining eight with ACIS-I (Table 1), all having the aimpoint within $\sim$1$\arcmin$ from the M87 nucleus, ensuring an optimal point-spread function (PSF). 
The total exposure is 677 ks.
The first three ACIS-S observations (ObsIDs 1808, 2707 and 3717) were used by Jord\'{a}n et al.~(2004) to study X-ray sources in the core region.
It is noteworthy that there exist tens of ACIS-S observations taken in the sub-array mode, primarily to study the X-ray variability of the M87 jet (Harris et al.~2003). 
We did not include these observations due to their small field-of-view.
We also neglected a 1-ks snapshot ACIS-I observation taken in 2000.


We used CIAO v4.8 and the calibration database CALDB v4.7.1 to reduce the data, following the standard procedure\footnote{http://cxc.harvard.edu/ciao}. 
We used only data from the S3 CCD for the ACIS-S observations and only data from I0-I3 for the ACIS-I observations, to ensure optimal sensitivity for source detection. 
We employed the CIAO tool {\it reproject\_aspect} to calibrate the relative astrometry among the individual observations, by matching the centroids of commonly-detected point sources.
For each observation, we produced counts and exposure maps in three bands (``Soft": 0.5--2 keV, ``Hard": 2--8 keV, ``Broad": 0.5--8 keV).  
The exposure maps were weighted by an absorbed power-law with a photon-index of 1.7 and a Galactic foreground absorption column density $N_{\rm H} = 2.0 \times 10^{20}{\rm~cm^{-2}}$.
We masked the ``readout streak" due to the bright nucleus of M87 in the counts map and the corresponding regions in the exposure map. 
The individual counts maps and exposure maps were then reprojected to a common tangential point, i.e., the nucleus of M87, to form the combined counts maps and exposure maps. 
We also created PSF maps for each observation and obtained an exposure-weighted average PSF map for source detection.
An exposure map-corrected 0.5--8 keV flux image of M87 is shown in Figure~2.

In addition to the 12 observations that together define the ``Center-Field", we also utilize 10 ACIS-I observations pointed toward a field (hereafter referred to as the "Off-Field") $\sim$20$\arcmin$ northwest to the nucleus of M87, with a total exposure of $\sim$460 ks (Table~1). 
The Off-Field is mainly used for probing X-ray sources in the halo of M87. 
Data of the Off-Field observations were reprocessed following the same procedure as for the Center-Field.  

We have examined the light curve of each ObsID and found that the particle background was quiescent during the vast majority of time intervals; only mild flares were present in ObsIDs 2707 and 3717. 
Hence we decided to preserve all the science exposures for the following source analysis, maximizing the available exposure time for source detection and characterization. 

\begin{table}[!h]
\centering
\linespread{1}
{
\smallskip
\caption{Log of {\it Chandra} observations}
\scriptsize
\begin{tabular}{llllll}
\hline
ObsID& RA &DEC & Obs Date & Exp\,(ks)  & CCDs \\
\hline
Center \\
\hline
1808          &12h30m49.4s 	&$+12^\circ23^\prime28^{\prime\prime}$& 2000-07-30&12.85     &  S3  \\ 
2707          &12h30m49.4s 	&$+12^\circ23^\prime28^{\prime\prime}$& 2002-07-06&98.66     &  S3  \\ 
3717          &12h30m49.4s 	&$+12^\circ23^\prime28^{\prime\prime}$& 2002-07-07&20.56     &  S3  \\ 
4007          &12h30m31.8s 	&$+12^\circ29^\prime26^{\prime\prime}$& 2003-11-21&36.18     &  S3  \\
5826          &12h30m49.5s 	&$+12^\circ23^\prime28^{\prime\prime}$& 2005-03-03&126.76    &  I0-I3  \\ 
5827          &12h30m49.5s 	&$+12^\circ23^\prime28^{\prime\prime}$& 2005-05-05&156.2     &  I0-I3  \\ 
5828          &12h30m49.5s 	&$+12^\circ23^\prime28^{\prime\prime}$& 2005-11-01&32.99     &  I0-I3  \\ 
6186          &12h30m49.5s 	&$+12^\circ23^\prime28^{\prime\prime}$& 2005-01-31&51.55     &  I0-I3  \\
7210          &12h30m49.5s 	&$+12^\circ23^\prime28^{\prime\prime}$& 2005-11-16&30.71     &  I0-I3  \\
7211          &12h30m49.5s 	&$+12^\circ23^\prime28^{\prime\prime}$& 2005-11-16&16.62     &  I0-I3  \\
7212          &12h30m49.5s 	&$+12^\circ23^\prime28^{\prime\prime}$& 2005-11-14&65.25     &  I0-I3  \\
11783         &12h30m57.7s 	&$+12^\circ16^\prime13^{\prime\prime}$& 2010-04-13&28.68     &  I0-I3  \\
\hline
Off \\
\hline
15178          &12h30m05.0s 	&$+12^\circ39^\prime53^{\prime\prime}$& 2014-02-17&46.48     &  I0-I3 \\
15179	      &12h30m05.0s 	&$+12^\circ39^\prime53^{\prime\prime}$& 2014-02-24&41.40     &  I0-I3  \\
15180	      &12h30m11.3s 	&$+12^\circ40^\prime24^{\prime\prime}$& 2013-08-01&138.77    &  I0-I3  \\
16585	      &12h30m05.0s 	&$+12^\circ39^\prime53^{\prime\prime}$& 2014-02-19&45.03     &  I0-I3  \\
16586	      &12h30m05.0s 	&$+12^\circ39^\prime53^{\prime\prime}$& 2014-02-20&49.19     &  I0-I3  \\
16587	      &12h30m05.0s 	&$+12^\circ39^\prime53^{\prime\prime}$& 2014-02-22&37.35     &  I0-I3  \\
16590	      &12h30m05.0s 	&$+12^\circ39^\prime53^{\prime\prime}$& 2014-02-27&37.59     &  I0-I3  \\
16591	      &12h30m05.0s 	&$+12^\circ39^\prime53^{\prime\prime}$& 2014-02-27&23.48     &  I0-I3  \\
16592	      &12h30m05.0s 	&$+12^\circ39^\prime53^{\prime\prime}$& 2014-03-01&35.61     &  I0-I3  \\
16593	      &12h30m05.0s 	&$+12^\circ39^\prime53^{\prime\prime}$& 2014-03-02&37.59     &  I0-I3  \\	 
\hline
\end{tabular}

}
\end{table}

\begin{figure}[h!]
\epsscale{1.0}
\includegraphics[width=3.5in]{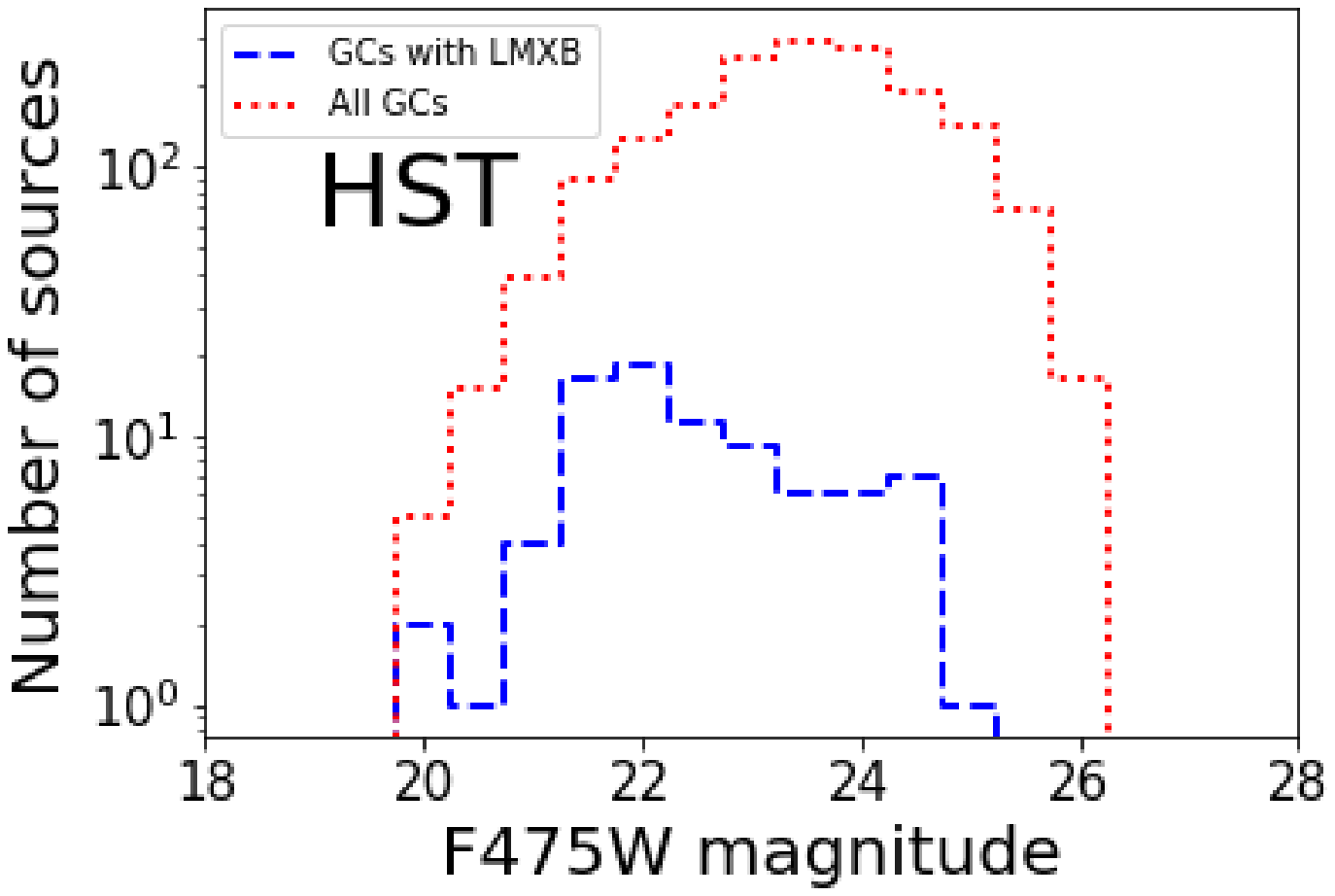}
\includegraphics[width=3.5in]{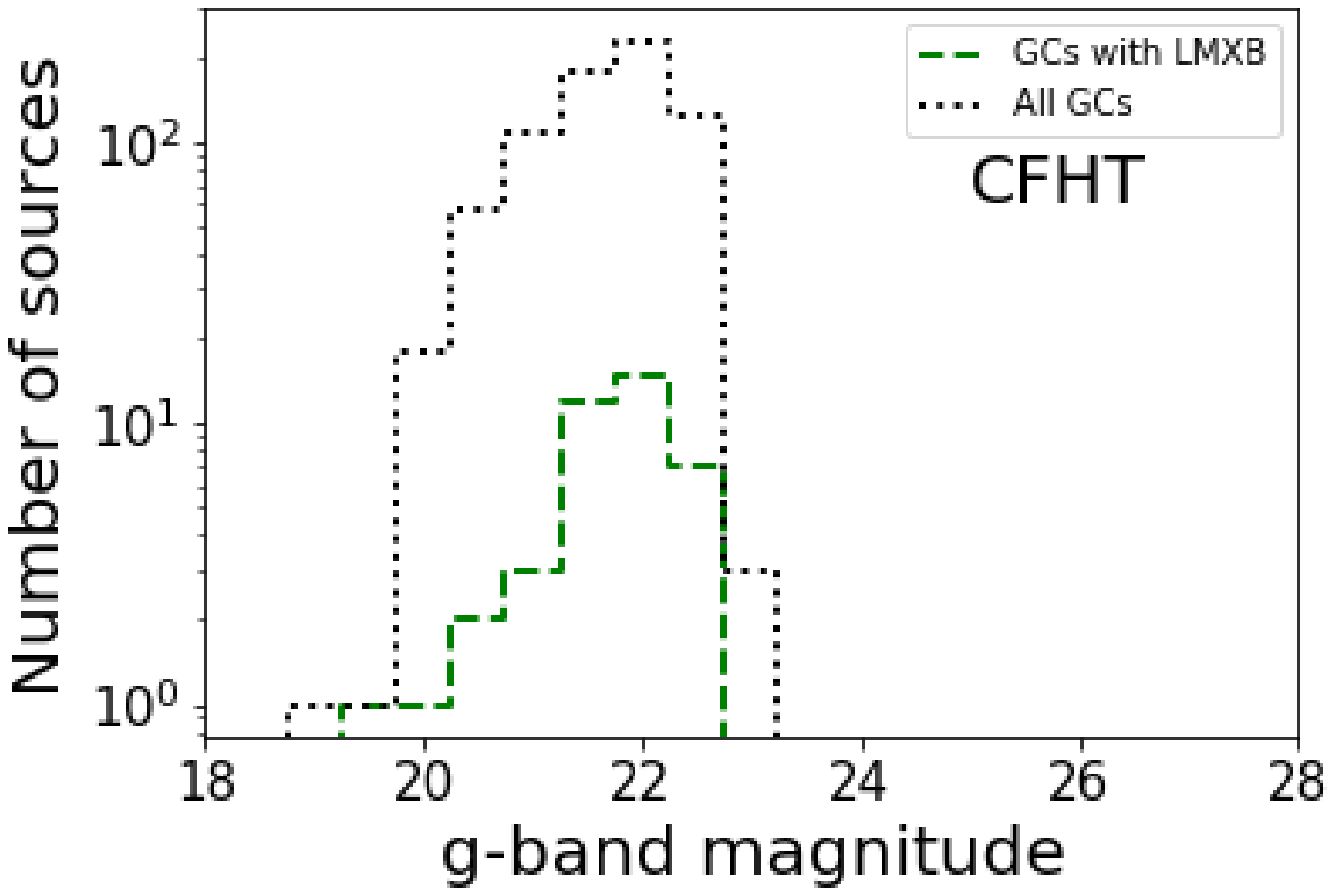}
\caption{Magnitude distributions of HST/ACSVCS GCs ({\it upper panel}) and CFHT/NGVCS GCs ({\it lower panel}). Those GCs with an X-ray counterpart are shown by the lower histogram.}
\label{fig:GC}
\end{figure}

\subsection{Globular Cluster Catalogs}
We utilized two publicly available catalogs of M87 GCs. 
The first is from the ACS Virgo Cluster Survey (ACSVCS) by HST (\citealp{Jordan2009}; hereafter called the ACSVCS catalog), which contains 1745 GC candidates. Among these, we selected 1668 objects with a $>$90\% probability of being a GC. We adopted the F475W and F850LP magnitudes from the Jord\'{a}n et al. catalog, which, due to similarity in the filter response, can be effectively taken as a $g$-band and a $z$-band magnitude, respectively. 
The ACSVCS catalog is complete down to the so-called GC turnover magnitude ($\sim$24.0 mag at $g$-band; Harris 2001), but it only covers a $\sim$10 arcmin$^{2}$ region in the core of M87. The exact ACSVCS footprint was described in Jord\'{a}n et al. (2004).
Therefore, we also employed the catalog from the Next Generation Virgo Cluster Survey (NGVCS) by CFHT (Powalka et al. 2016; hereafter called the NGVCS catalog) to identify M87 GCs located at large galactocentric radii.  
A total of 851 GCs from this catalog were selected, which are not in the ACSVCS catalog, but are within either the Center-Field or the Off-Field.
We note that the NGVCS catalog is not sensitive to finding GCs with a $g$-band magnitude fainter than 23 mag and thus is largely incomplete below the turnover magnitude (Figure~3).
We estimate that at least $\sim$2900 GCs were not detected in NGVCS, assuming the same intrinsic GC magnitude distribution in the ACSVCS and NGVCS footprints.

\begin{table*}
\label{table:center_source}
\centering
\linespread{1.2}
{\small

\smallskip
\caption{X-ray Point Sources in the Center-Field}
\begin{tabular}{cccccccccc}
\hline\noalign{\smallskip}
\hline\noalign{\smallskip}
\\

 ID& RA &Dec & positional error& $S_{0.5-2.0}$  & $S_{2.0-8.0}$& $S_{0.5-8.0}$ & $F_{0.5-8.0}$  &$S_{\rm max}/S_{\rm min}$ &Note\\
(1) & (2) & (3) & (4) & (5) & (6) & (7) &(8)&(9) & (10)\\

\hline\noalign{\smallskip}
1	&	12:30:08.43	&	12:23:28.14	&	0.54	&$	125.5	^{+	9.9	}_{-	9.9	}$&$	9.3	^{+	3.1	}_{-	3.1	}$&$	95.7	^{+	6.8	}_{-	7.6	}$&$	28.0	$&		&			\\
2	&	12:30:14.13	&	12:23:47.62	&	0.59	&$	<28.2	$&$	17.9	^{+	3.3	}_{-	3.0	}$&$	29.0	^{+	4.3	}_{-	4.9	}$&$	8.5	$&		&	G		\\
3	&	12:30:15.09	&	12:27:55.61	&	0.73	&$	<41.6	$&$	15.8	^{+	3.5	}_{-	3.4	}$&$	33.2	^{+	5.3	}_{-	6.0	}$&$	9.7	$&		&			\\
4	&	12:30:17.72	&	12:25:45.58	&	0.29	&$	112.9	^{+	6.3	}_{-	6.3	}$&$	6.8	^{+	2.0	}_{-	2.0	}$&$	106.7	^{+	4.7	}_{-	5.0	}$&$	31.3	$&	1.54	&		V	\\
5	&	12:30:17.73	&	12:19:47.23	&	0.59	&$	22.4	^{+	5.6	}_{-	5.4	}$&$	3.4	^{+	1.1	}_{-	1.2	}$&$	12.3	^{+	3.3	}_{-	3.2	}$&$	3.6	$&		&			\\
6	&	12:30:17.99	&	12:26:36.39	&	0.39	&$	35.5	^{+	4.9	}_{-	4.7	}$&$	10.0	^{+	2.7	}_{-	2.5	}$&$	25.7	^{+	3.0	}_{-	3.1	}$&$	7.5	$&		&			\\
7	&	12:30:18.06	&	12:19:22.95	&	0.40	&$	93.3	^{+	8.0	}_{-	8.4	}$&$	18.2	^{+	2.7	}_{-	2.5	}$&$	54.4	^{+	4.9	}_{-	5.1	}$&$	15.9	$&		&			\\
8	&	12:30:19.87	&	12:18:39.41	&	0.44	&$	31.5	^{+	6.8	}_{-	6.5	}$&$	9.1	^{+	2.4	}_{-	2.2	}$&$	32.4	^{+	3.9	}_{-	4.4	}$&$	9.5	$&		&			\\
9	&	12:30:19.91	&	12:19:21.16	&	0.66	&$	<1.9	$&$	10.6	^{+	1.9	}_{-	1.8	}$&$	9.3	^{+	1.8	}_{-	2.2	}$&$	2.7	$&		&			\\
10	&	12:30:20.54	&	12:26:47.31	&	0.35	&$	25.1	^{+	4.0	}_{-	3.9	}$&$	31.1	^{+	2.3	}_{-	2.3	}$&$	22.5	^{+	2.8	}_{-	2.9	}$&$	6.6	$&		&			\\

\hline
\end{tabular}
\tablecomments{(1) Source ID; (2)-(3) RA and DEC of source centroid in epoch J2000; (4) Positional error estimated by net counts and off-axis angle in units of arcsecond; (5)-(7) 0.5-2, 2-8 and 0.5-8 keV photon fluxes and errors, in units of $\rm 10^{-7}\ photon~cm^{-2}~s^{-1}$. For sources undetected in a certain band for $\sigma$, a $3\sigma$ upper limit is provided;
(8) 0.5-8 keV unabsorbed energy flux, in units of $\rm 10^{-15}\ erg~cm^{-2}~s^{-1}$;
(9) Ratio of maximum flux to minimum flux as measured among the individual observations, shown only for sources with significant variability;
(10) `G' denotes sources positionally coincident with an optically-identified GC; `V' denotes sources with strong variability. 
(Only a portion of the full table is shown here to illustrate its form and content.)}
}

\end{table*}

\section{X-ray Point Source Catalog}
\begin{table*}
\label{table:off_source}
\centering
\linespread{1.2}
{\small

\smallskip
\caption{X-ray Point Sources in the Off-Field}

\begin{tabular}{ccccccccc}
\hline\noalign{\smallskip}
\hline\noalign{\smallskip}
\\

 ID& RA &Dec & positional error& $S_{0.5-2.0}$  & $S_{2.0-8.0}$& $S_{0.5-8.0}$ & $F_{0.5-8.0}$ &Note  \\
(1) & (2) & (3) & (4) & (5) & (6) & (7) &(8)&(9)\\

\hline\noalign{\smallskip}
1	&	12:29:24.98	&	12:41:31.38	&	0.62	&$	92.9	^{+	10.3	}_{-	9.8	}$&$	109.7	^{+	7.0	}_{-	7.0	}$&$	239.2	^{+	12.7	}_{-	13.6	}$&	83.9	&		\\
2	&	12:29:32.52	&	12:38:48.07	&	1.45	&$	<3.7$&$	<3.5$&$	2.0	^{+	0.5	}_{-	1.9	}$&	0.7	&		\\
3	&	12:29:34.15	&	12:38:11.52	&	1.19	&$	12.2	^{+	5.7	}_{-	5.4	}$&$	<13.4$&$	15.2	^{+	5.4	}_{-	6.1	}$&	5.3	&		\\
4	&	12:29:34.53	&	12:48:48.52	&	1.51	&$	<32.4$&$	<8.6$&$	9.7	^{+	2.4	}_{-	9.5	}$&	3.4	&		\\
5	&	12:29:34.99	&	12:46:28.75	&	1.30	&$	<17.6$&$	14.6	^{+	3.4	}_{-	3.2	}$&$	23.3	^{+	4.7	}_{-	5.6	}$&	8.2	&		\\
6	&	12:29:36.24	&	12:36:35.18	&	0.80	&$	30.6	^{+	6.9	}_{-	6.9	}$&$	16.6	^{+	4.2	}_{-	3.8	}$&$	48.8	^{+	7.8	}_{-	7.8	}$&	17.1	&		\\
7	&	12:29:37.23	&	12:41:32.72	&	0.79	&$	20.4	^{+	4.5	}_{-	4.1	}$&$	<7.3$&$	21.0	^{+	4.1	}_{-	4.5	}$&	7.4	&		\\
8	&	12:29:38.86	&	12:47:53.22	&	1.09	&$	26.0	^{+	5.8	}_{-	5.5	}$&$	11.1	^{+	3.8	}_{-	3.4	}$&$	37.6	^{+	6.6	}_{-	7.1	}$&	13.2	&		\\
9	&	12:29:39.13	&	12:44:16.66	&	0.90	&$	<18.8$&$	20.0	^{+	5.6	}_{-	5.8	}$&$	28.9	^{+	7.1	}_{-	8.7	}$&	10.1	&		\\
10	&	12:29:39.14	&	12:50:25.31	&	3.26	&$	<15.6$&$	14.2	^{+	4.4	}_{-	4.0	}$&$	22.5	^{+	5.9	}_{-	6.6	}$&	7.9	&		\\

\hline
\end{tabular}
\tablecomments{(1) Source ID; (2)-(3) RA and DEC of source centroid in epoch J2000; (4) Positional error estimated by net counts and off-axis angle in units of arcsecond; (5)-(7) 0.5-2, 2-8 and 0.5-8 keV photon fluxes and errors, in units of $\rm 10^{-7}\ photon~cm^{-2}~s^{-1}$. For sources undetected in a certain band for $\sigma$, a $3\sigma$ upper limit is provided;
(8) 0.5-8 keV unabsorbed energy flux, in units of $\rm 10^{-15}\ erg~cm^{-2}~s^{-1}$;
(9) `G' denotes sources positionally coincident with an optically-identified GC.
(Only a portion of the full table is shown here to illustrate its form and content.)}
}

\end{table*}

We used the CIAO tool {\it wavdetect} to detect point-like sources in the three energy bands (Soft: 0.5-2.0 keV; Hard: 2.0-8.0 keV and Broad: 0.5-8.0 keV), with a local false detection probability $P \leq 10^{-6}$. 
A total of 473 sources were detected in the soft band, 289 sources in the hard band and 564 sources in the broad band. 
We combined these detections, omitting duplicate sources (defined to be positionally coincident within errors), to obtain a list of 645 independent sources. 
The X-ray-bright and extended jets in M87, as well as ``clumps" of dense, hot gas in the core region (outlined by the polygon in Figure~2), can mimic compact sources that inevitably picked up by {\it wavdetect} or any other source detection algorithm tuned to search for ``local peaks".
A similar situation was encountered, for instance, in the case of NGC\,5128 (\citealp{Voss2009}) and the Galactic center (\citealp{Muno2009}; \citealp{Zhu2018}).
Similar to the approach in these work, we visually identified such spurious sources, which exclusively lie within the polygon region as outlined in Figure~2. 
A total of 299 spurious sources were thus removed from the raw source list.
For reference, we provide the approximate positions of these spurious sources in the Appendix.
The remaining 346 sources are considered truly point-like, the positions of which are shown in Figure~2. 
Using the CIAO tool {\it lim\_sens}, we determined the detection sensitivity across the field-of-view. 
The median sensitivity within the inner $4\arcmin$ region is found to be $4.3 \times10^{-7}{\rm~photon~cm^{-2}~s^{-1}}$. 

The initial source centroids obtained from {\it wavdetect} were improved by a maximum likelihood method that iterates over source counts detected within the exposure-weighted mean 90\% enclosed-energy radius (EER). 
Aperture photometry was then performed to obtain the source photon fluxes. 
In each observation, source counts were extracted from within the 90\% EER as determined by the individual PSF map, while background region was defined for each source by taking an annulus typically with 2-3 times the 90\% EER, excluding any pixels falling with 2 times the 90\% EER of the neighboring sources as well as pixels affected by the spurious sources that have already been removed.
The net source counts of a give source were then summed up from the individual observations, and the photon fluxes were derived by dividing the total effective exposure at the source position.
We calculated the errors of the photon fluxes in the three bands using a Byasian algorithm (\citealp{Park2006}). 
The observed photon flux was converted to an unabsorbed energy flux using a factor of $2.93\times10^{-9}{\rm~erg~photon^{-1}}$, again assuming the fiducial incident spectrum over the energy range of 0.5--8.0 keV.

A similar procedure was performed for the Off-Field, in which 103 point sources were detected in the soft band, 124 sources in the hard band and 157 sources in the broad band.
A total of 173 independent sources are resulted after comparing the source positions in the three bands.  We found no need to filter diffuse ``clumps" in the Off-Field.
The median sensitivity within 4$\arcmin$ from the geometric center of the Off-Field, [RA, DEC]=[12:30:07.6,+12:41:30.5], is $2.3\times10^{-7}{\rm~photon~cm^{-2}~s^{-1}}$,
which is actually better than in the Center-Field due to the much lower diffuse background.

The source positional uncertainties ($PU$), at 68\% confidence level, were estimated following the empirical relation of \citet{Kim2007}, based on simulations, 
\begin{eqnarray}
\tiny
 {log PU} = \left\{
\begin{array}{ll}
      0.1137OAA-0.4600logC-0.2398,    &log C \leq 2.1227 \\
      0.1031OAA-0.1945logC-0.8034,    &  2.1227 < log C \leq 3.300 \\
\end{array} 
\right.
\end{eqnarray}
where $PU$ is in units of arcseconds and $OAA$ is the off-axis angle in arcminutes. For each source, we calculated the value of $OAA$ as the exposure-weighted mean of $OAA$ in the individual observations.
We found that the average $PU$ is $\sim$0\farcs1 for sources located within $4^\prime$ from the M87 nucleus, and $\sim$0\farcs3 for sources detected further outside.

A catalog of the point sources detected in the Center-Field and Off-Field is given in Table 2 and Table 3, respectively, which includes information on source centroids, positional uncertainties, photon fluxes in the three bands, and the 0.5--8 keV unabsorbed flux.   

\section{Analysis and results}
\subsection{X-ray sources associated with GCs} {\label{subsec:XGC}

We cross-correlated the detected X-ray sources with the two GC catalogs, following the method of \citet{Hong2009}. 
Among the 346 sources detected in the Center-Field, 160 fall within the ACSVCS footprint, and all are covered by the NGVCS. For the Off-Field, all 173 sources are covered by the NGVCS.
We did not consider any NGVCS GCs falling within the ACSVCS footprint, because the ACSVCS catalog should be more accurate and more complete than the NGVCS catalog. 
For a given X-ray source and its closest GC, their relative distance, $d_{\rm r}$, is defined as the ratio of the angular offset to the quadratic sum of their positional uncertainties. 
The distribution of $d_{\rm r}$ is naturally a bimodal, with the first peak attributed to true X-ray counterparts of the GCs, and the second peak dictated by the source surface density. 
We adopted the local minimum between the two peaks, $d_{\rm r} \leq 10.0$, as a natural cut to select genuine associations between an X-ray source and a GC. 
This results in a total of 122 matches in the Center-Field, among which 81 matches are found for the ACSVCS catalog and 41 for the NGVCS catalog. 
For comparison, \citet{Jordan2004} found 58 X-ray sources associated with an ACSVCS GC. 
We also tested the random match probability by artificially shifting the GC positions by an amount of 10$^{\prime\prime}$. In this way, we found 5 false matches, i.e., less than 5\% of the 122 pairs might be random matches, among which 4 occurred for the ACSVCS GCs and 1 for the NGVCS GCs.
The 122 X-ray sources associated with GCs are denoted by `G' in Table 2. 

Among all the GCs within the {\it Chandra} field-of-view, 4.9$\pm$0.5\% of the ACSVCS GCs and 5.9$\pm$0.9\% of the NGVCS GCs have an X-ray counterpart. On the other hand, the percentage of X-ray sources found in the ACSVCS GCs and NGVCS GCs outside the HST footprint is 50.6$\pm$5.6\% and 22.0$\pm$3.4\%, respectively. 
The difference could be due partly to the greater incompleteness in the NGVCS GCs and partly to the increasing contamination of cosmic X-ray background (CXB) sources (primarily distant AGNs) at large galactocentric radii. 
If the CXB contamination were statistically subtracted (see Section 4.3), the percentage becomes $52.8\%\pm5.9\%$ for the ACSVCS GCs and $50.9\%\pm7.9\%$ for NGVCS GCs, respectively.
In the Off-Field, we found 6 GCs having an X-ray counterpart, or 4.6$\pm$1.8\% of the NGVCS GCs in this field, which are denoted by `G' in Table 3.
 
\begin{figure}[h!]
\epsscale{1.0}
\includegraphics[width=3.5in]{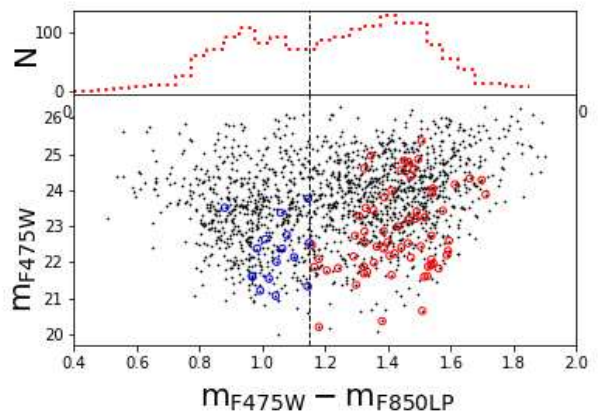}
\caption{\small Color-magnitude diagram of the HST/ACSVCS GCs. Those GCs with an X-ray counterpart are marked by an open circle. 
The top panel shows the color histogram of all HST/ACSVCS GCs.
The vertical dashed line, at a color of 1.15, marks the adopted division of blue and red GCs.}
\label{fig:m_c}
\end{figure}


Figure~\ref{fig:m_c} shows the color-magnitude diagram of the ACSVCS GCs.
We adopted a threshold of $m_{\rm F475W}-m_{\rm F850LP} = 1.15$ to divide blue and red GCs.  Due to the small difference between the $g-z$ color and $m_{\rm F475W}-m_{\rm F850LP}$ color, we also used $g-z = 1.15$ to distinguish the red and blue GCs in the NGVCS catalog.
Those GCs with an X-ray counterpart are further marked by an open circle in Figure~\ref{fig:m_c}.
Among the 81 ACSVCS GCs with an X-ray counterpart, 16 are blue and 65 are red. This corresponds to 2.5$\pm$0.6\% of all blue GCs, and 5.6$\pm$0.8\% of all red GCs, to host an X-ray source. 
That the red (metal-rich) GCs have $\sim$2.2 times higher probability than the blue (metal-poor) GCs to host an X-ray source is broadly consistent with previous work on ETGs (e.g.,  \citealp{Fabbiano2006}). 

In addition to the GCs, ultra-compact dwarfs (UCDs) were also searched for an X-ray counterpart.
We utilized the UCD catalog of \citet{Ko2017}, which includes 52 UCDs within the Center-Field and 24 UCDs in the Off-Field.
However, we found no positional matches between these UCDs and our X-ray sources.

\subsection{Hardness ratios and cumulative spectra}

It was impractical to constrain the spectrum of each detected source due to the limited source counts from most sources. 
Instead, we examined the hardness ratio between the soft-band (0.5--2.0 keV) and the hard-band (2.0--8.0 keV), defined as,  
\begin{eqnarray}
HR \equiv {(C_{\rm H}-C_{\rm S})\over(C_{\rm H}+C_{\rm S})},
\end{eqnarray}
where $C_{\rm H}$ and $C_{\rm S}$ are the net source counts. Errors in $HR$ are estimated using a Bayesian algorithm (\citealp{Park2006}). 
Figure~\ref{fig:xcmd} shows the hardness ratio versus the 0.5--8 keV unabsorbed luminosity, in comparison with predicted hardness ratios of certain absorbed power-laws. The bulk of sources exhibit a hardness ratio consistent with LMXBs and/or background AGNs, i.e., with photon-indices of 1--2. About 20\% sources exhibit either a hard or soft color that apparently lie beyond this typical range. Most of these sources have a relatively low flux and a large uncertainty in the hardness ratio. The hard sources might be heavily obscurced AGNs; 12 of the soft sources are associated with a GC, while the rest might be background galaxies of strong star formation. 
This is supported by the finding of a similar fraction of hard/soft sources in the Off-Field, in which the vast majority of detected sources should be background sources (Section 4.3).

\begin{figure}[h!]
\epsscale{1.0}
\includegraphics[width=0.5\textwidth]{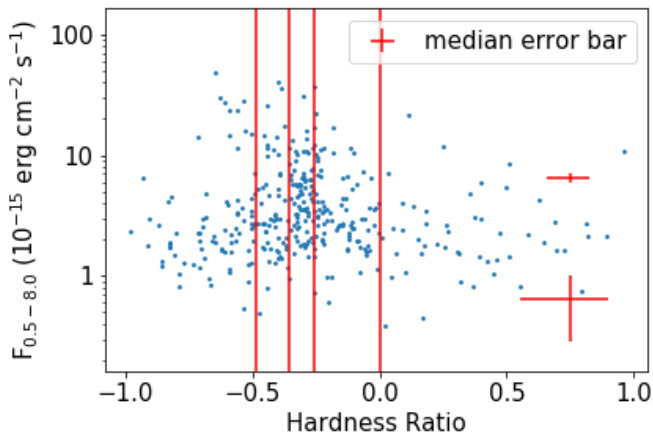}
\caption{Hardness ratio versus X-ray luminosity. Ideal absorbed power-law spectra are shown as red vertical lines, with the Galactic foreground absorption and photon-indices of 2.0, 1.7, 1.5 and 1.0 from left to right. The upper (lower) error bar on the right illustrates the median uncertainty in the hardness ratio and unabsorbed flux of sources brighter (fainter) than $3\times10^{-15}{\rm~erg~cm^{-2}~s^{-1}}$.}
\label{fig:xcmd}
\end{figure}

\begin{figure}
\epsscale{1.0}
\includegraphics[width=2.5in,angle=270]{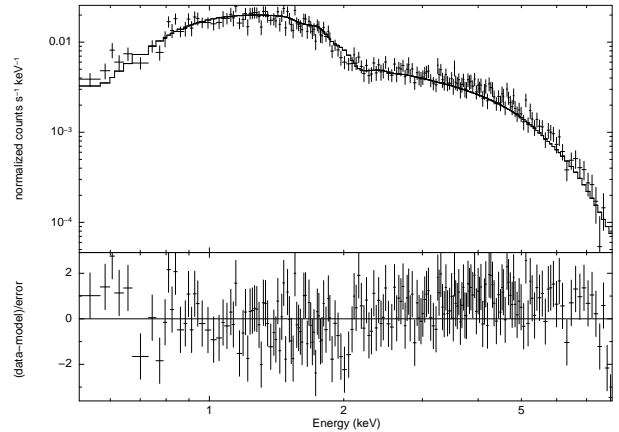}

\caption{The cumulative spectrum of GC-LMXBs. Also plotted are the best-fit absorbed bremsstrahlung models.}
\label{fig:spec}
\end{figure}

We also examined the cumulative spectrum 
extracted from the 122 sources associated with GCs (hereafter GC-LMXBs; Figure~\ref{fig:spec}). 
The source and background spectra were extracted using regions same as for the photometry (Section 3).
The ancillary response files (ARFs) and redistribution matrix files (RMFs) were obtained by weighting the effective exposure at the individual source positions.
Since only the eight ACIS-I observations have a sufficient field-of-view to cover all these sources, we did not include the four ACIS-S observations for the spectral analysis. 


We used XSPEC v12.9.0 to fit the background-subtracted source spectrum over the energy range of 0.5--8 keV.
We tested a power-law model and a bremsstrahlung model, both subject to a free absorption. 
For the absorbed power-law, the best-fit photon-index is $1.84\pm0.03$, with $\chi^{2}$/degree of freedom = 727.9/509.
For the absorbed bremsstrahlung model, the best-fit plasma temperature is $6.18\pm0.38$ keV, with $\chi^{2}$/degree of freedom=656.6/509. 
In both cases, the absorption column density is $\sim$ $9\times10^{20}{\rm~cm^{-2}}$.


\subsection{Luminosity function}

\begin{figure}
\epsscale{1.0}
\includegraphics[width=3.5in]{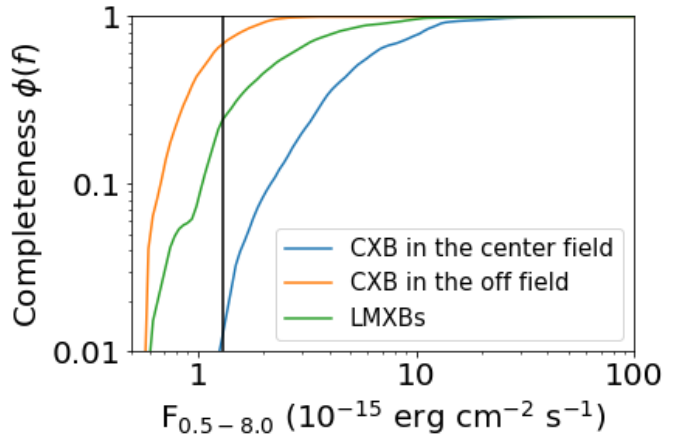}
\caption{\small The completeness function of the LMXBs and CXB. 
The vertical line marks the flux value (corresponding to a luminosity of $4\times10^{37}{\rm~erg~s^{-1}}$) above which the luminosity functions are fitted (Section 4.3).} 
\label{fig:complete}
\end{figure}

\begin{figure*}
\epsscale{1.0}
\includegraphics[width=3.5in]{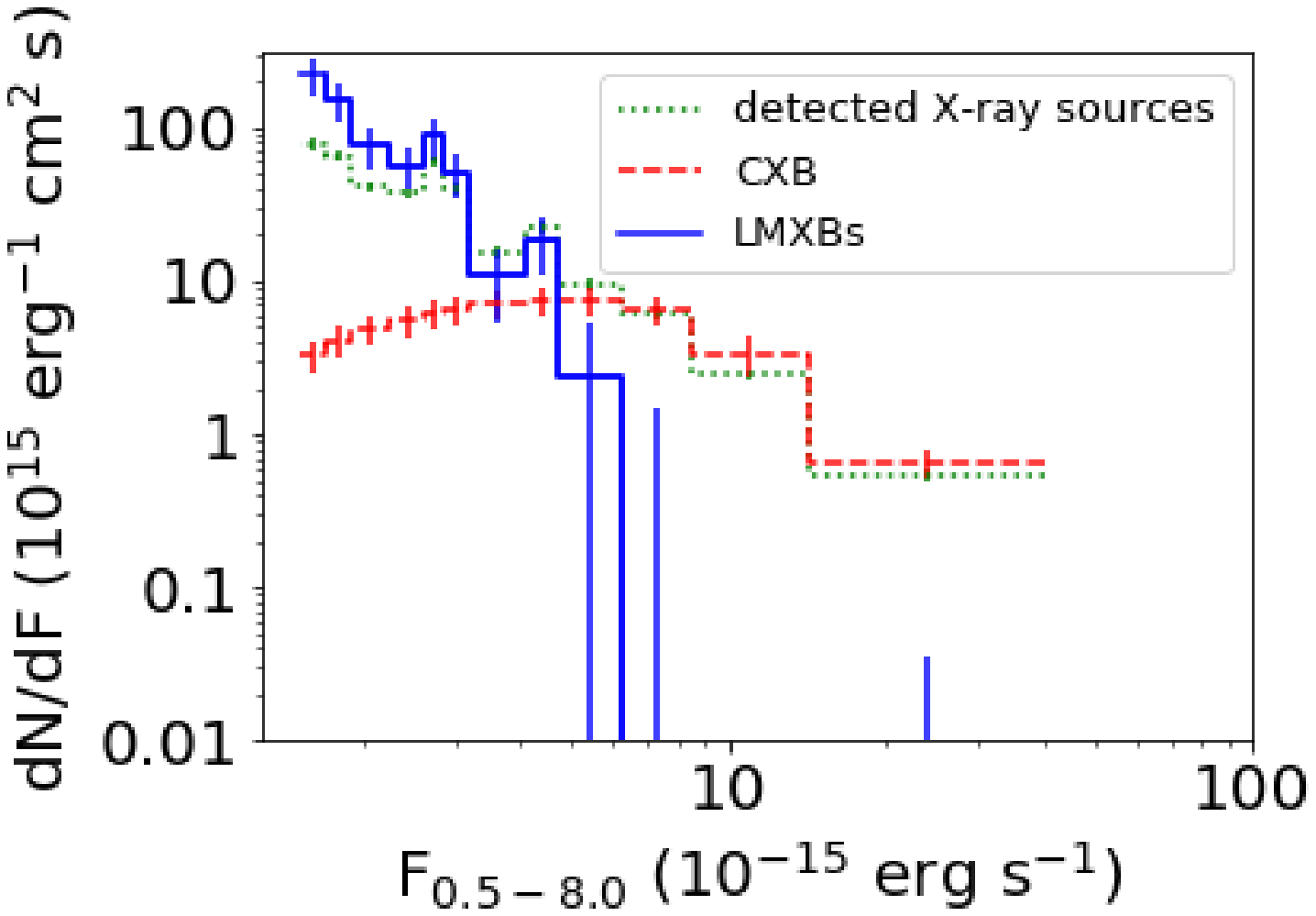}
\includegraphics[width=3.5in]{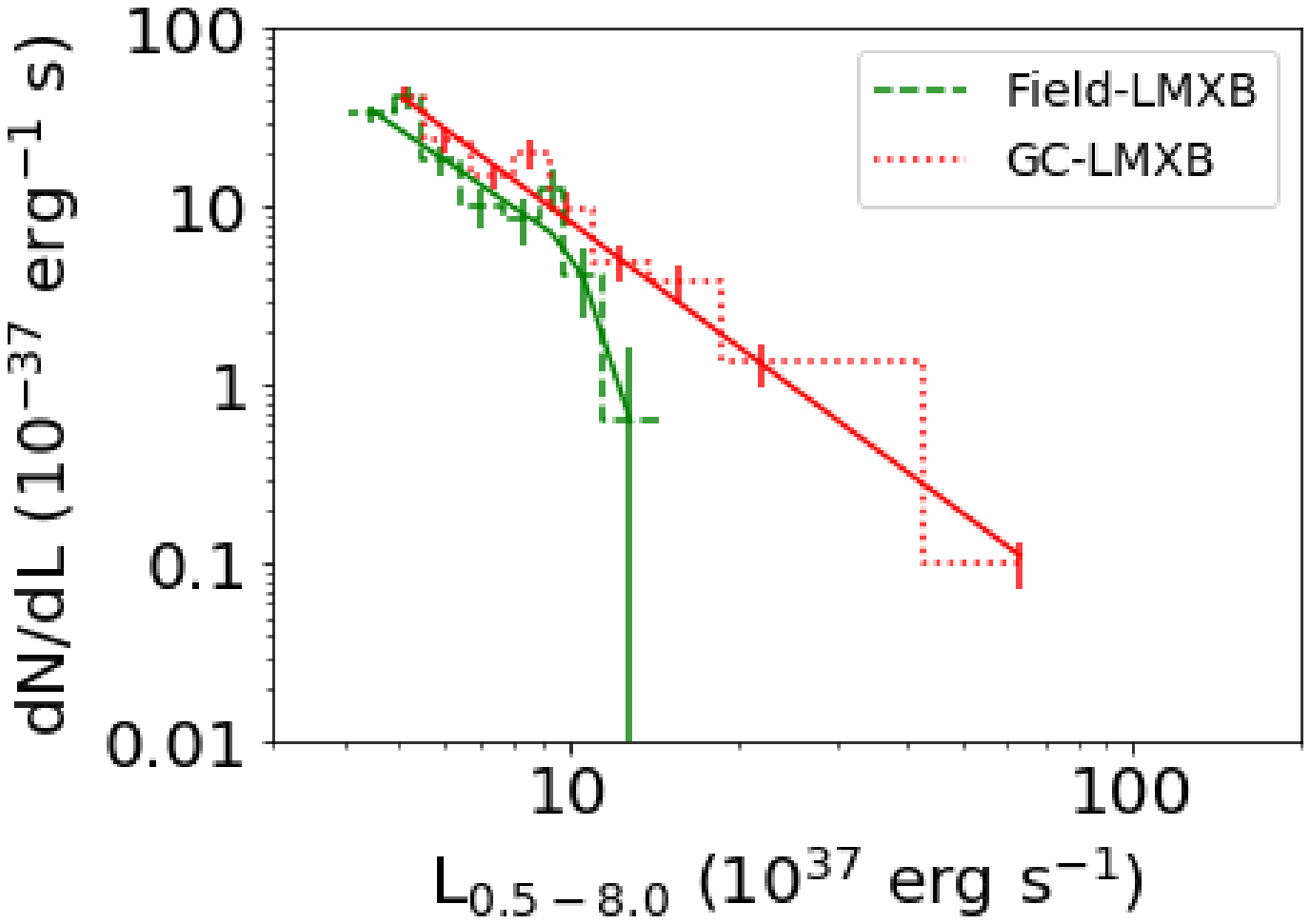}
\caption{\small {\sl Left}: The observed luminosity function of all detected sources is shown by the green histogram. The predicted CXB contribution, taken in account the detection sensitivity at the faint end, is shown by the red curve. The difference, i.e., representing the LMXB-only luminosity function, is shown by the blue histogram. 
{\sl Right}: Luminosity functions of all LMXBs (blue), GC-LMXBs (red) and field-LMXBs (green), corrected for the incompleteness (Figure 7). Also shown are the best-fit broken power-law for field-LMXBs and the single power-law for GC-LMXBs (Table 4).
}
\label{fig:LF}
\end{figure*}

To determine the intrinsic luminosity function of the detected X-ray sources, we need to first account for the variation in the detection sensitivity across the field-of-view (Section 3). 
The detection completeness $\phi(f)$, defined as the fraction of sources at a given flux $f$ that would be detected in the observations, was calculated following the method of \citet{Voss2006}.
Specifically, $\phi(f)$ was computed from the fraction of pixels with a sensitivity better than $f$, weighted by the source spatial distribution.
For both the field-LMXBs and GC-LMXBs, we assume that their spatial distribution follows the starlight distribution (\citealp{Gilfanov2004}), although GCs could have a somewhat broader radial distribution than the field stars.
Stellar mass distribution was estimated from the $I$-band radial profile (Cappellari et al. 2006). 
It is noteworthy that when calculating $\phi(f)$ and constructing the observed luminosity functions, we have excluded regions where spurious sources were identified (polygon in Figure~2), as well as the sources detected therein.

A significant fraction of the detected X-ray sources is expected to originate from the CXB. 
We estimated the number of CXB sources based on the empirical log$N$-log$S$ relation of \citet{Georgakakis2008}, which has a broken power-law form,  
\begin{eqnarray}
 {dN\over df} = \left\{
\begin{array}{ll}
      K{(f/f_{ref})}^{-\beta_{1}} &f < f_{b} \\
      K'{(f/f_{ref})}^{-\beta_{2}} &f \geq f_{b} \\
\end{array} 
\right.
\end{eqnarray}
where $f_{ref}=10^{-14}{\rm~erg~s^{-1}~cm^{-2}}$ and $K'= K{(f_{b}/f_{ref})}^{\beta_{1}-\beta_{2}}$. 
The adopted parameters are $\beta_{1}=-1.58$, $\beta_{2}=-2.48$, $f_{b}=2.63\times10^{-14}{\rm~erg~s^{-1}~cm^{-2}}$ and $K=3.74\times10^{16}{\rm~deg^{-2}}/({\rm erg~s^{-1}~cm^{-2}})$, which are suitable for CXB sources detected in the 0.5--8 keV band.
The completeness function of sources (mostly AGNs) associated with the CXB was calculated assuming that they were distributed uniformly across the field. The derived completeness functions are shown in Figure~\ref{fig:complete}. The number of CXB sources is estimated to be 112 in the Center-Field and 134 in the Off-Field, respectively. 
Following \citealp{Lahav1992} and \citealp{Li2010}, we estimated that the above values are subject to a cosmic variance of $\sim$20\%, which is to be added to the Poisson errors when calculating the uncertainties in the following luminosity functions. 


The observed luminosity functions, after subtraction of the CXB (for the case of field-LMXBs) and correction for the incompleteness, are shown in Figure~\ref{fig:LF}b GC-LMXBs and field-LMXBs, respectively.                
We included only sources detected in the 0.5--8 keV band to be consistent with the CXB estimate.
The GC-LMXBs exhibit a significantly flatter luminosity function than the field-LMXBs.
We fitted the luminosity functions of GC-LMXBs with a simple power-law,
\begin{equation}{dN\over dL_X} \propto 
\begin{array}{ll} L^{\alpha}_X,  \\
\end{array} 
\end{equation}
which results in $\alpha = 2.34\pm0.09$, with $\chi^2$/dof of 3.60/8. 
Since the incompleteness of the NGVCS GCs might affect the result, we also constructed and fitted a luminosity function using the ACSVCS GC-LMXBs only. 
This results in a somewhat steeper slope of $\alpha = 2.53\pm0.13$. 
We also tried a broken power-law fit,
\begin{equation}{dN\over dL_X} \propto \left\{
\begin{array}{ll}
      L^{\alpha_1}_X &  L_X < L_b \\
      L^{\alpha_2}_X & \ \ L_X \geq L_b  \\

\end{array} 
\right.
\end{equation}
which results in $\chi^2$/dof of 2.94/6 for all GC-LMXBs.
According to the F-test, the broken power-law is a better model for GC-LMXBs at only 46\% confidence.
Similarly, we fitted the luminosity function of the Field-LMXBs using both power-law and broken power-law, and the F-test suggets that the broken power-law is a better model at 58\% confidence.
The best-fit values of $\alpha$, $\alpha_1$, $\alpha_2$ and $L_b$ are given in Table 4. 

\begin{table}[!h]
\centering
\linespread{1.5}
{\small

\smallskip
\caption{Fitted Luminosity Functions}
\begin{tabular}{ccc}
\hline\noalign{\smallskip}
\hline\noalign{\smallskip}
Parameter                  & GC-LMXBs  & Field-LMXBs \\
\hline\noalign{\smallskip}
$\alpha$           &  2.34$\pm$0.09   &  3.33$\pm$0.70 \\
$\chi^2$/dof & 3.60/8 & 3.78/6 \\

\hline\noalign{\smallskip}
$\alpha_1$         &1.81$\pm$0.69     &  2.15$\pm$0.61  \\ 
$\alpha_2$              &  2.44$\pm$0.15   &  $9.6^{+10.0}_{-9.6}$  \\ 
$L_{b}\,(10^{37}{\rm~erg~s^{-1}})$                  &8.6$\pm$3.6     &  10.2$\pm$1.1  \\ 
$\chi^2$/dof & 2.94/6 & 2.80/4 \\

\hline\noalign{\smallskip}

\hline\noalign{\smallskip}

\end{tabular}

}
\end{table}

\subsection{Radial distribution}

\begin{figure*}
\epsscale{1.0}
\includegraphics[width=3.5in]{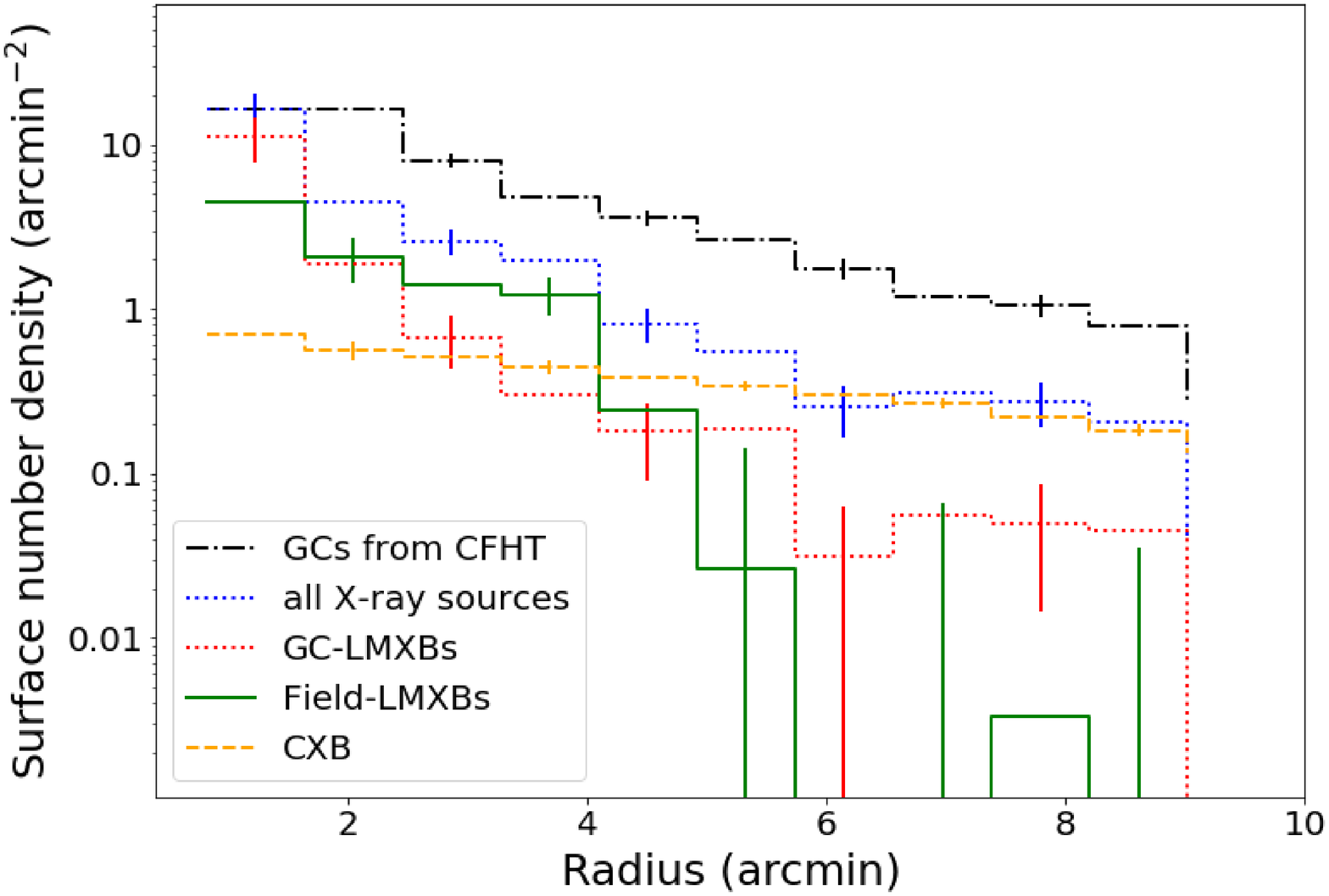}
\includegraphics[width=3.5in]{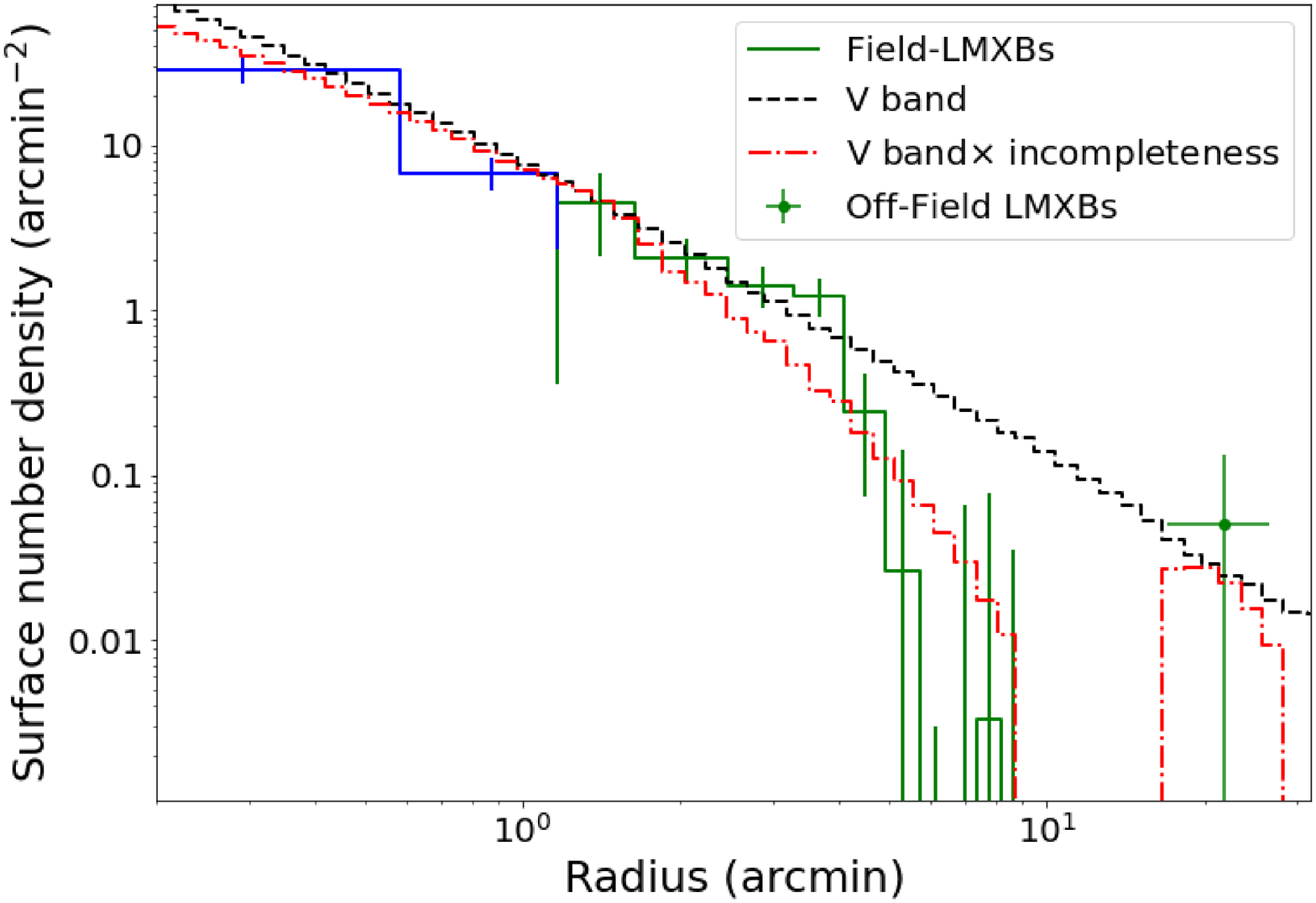}
\caption{\small Azimuthally-averaged source surface density profiles. (a) Blue: all detected X-ray sources; Red: GC-LMXBs; Green: field-LMXBs (i.e., CXB-subtracted). The predicted CXB profile, taking into account detection sensitivity, is shown by the orange curve. For comparison, the NGVCS-GC profile is shown by the black histogram. (b) The surface density profile of the field-LMXBs is compared with the $V$-band starlight profile, a proxy of the projected stellar mass distribution. The intrinsic $V$-band starlight profile is shown by the black curve, while the red curve takes into account the position-dependent detection sensitivity. The starlight profile is normalized by matching the first green data point.
The innermost two data points in blue represent sources found in the polygon region as shown in Figure 2, which might be subject to incompleteness. The outermost data point represents the mean surface density of X-ray sources detected in the Off-Field, after subtracting the predicted CXB contribution therein.
In both panels, only X-ray sources brighter than $4\times 10^{37}{\rm~erg~s^{-1}}$ are considered, to minimize the effect of detection incompleteness (Figure 7).
}
\label{fig:RP}
\end{figure*}

Next, we examined the azimuthally-averaged surface density profiles of the X-ray sources, as shown in Figure~\ref{fig:RP}. 
Only sources brighter than $1.3\times10^{-15}{\rm~erg~cm^{-2}~s^{-1}}$ ($\sim$ $4\times10^{37}{\rm~erg~s^{-1}}$ at the distance of M87) are considered, to minimize the effect of detection incompleteness.
We have also excluded sources found within the masked regions as outlined by the polygon in Figure 2. 
The surface density profile of all the remaining sources, corrected for the fractional coverage of a given annulus, is shown by the blue histogram, which presumably consists of LMXBs (field-LMXBs and GC-LMXBs) and CXB sources. 
We estimated the contribution from the CXB (orange curve), using the empirical log$N$-log$S$ relation (Section 4.3) and taking into account the position-dependent detection sensitivity (Section 3).
It can be seen that the CXB dominates the detected X-ray sources at galactocentric radii $\gtrsim6\arcmin$.   
After subtracting the CXB contribution, we obtained the LMXB component, which is divided into GC-LMXBs (red histogram) and field-LMXBs (green histogram). 
Clearly, the GC-LMXBs have a broader radial distribution than the field-LMXBs, which is consistent with the distribution of all NGVCS GCs shown as the black histogram, except in the inner $\sim$3$\arcmin$ where the sensitivity of NGVCS is strongly affected by the diffuse starlight of M87.  

As for the field-LMXBs, it is expected that they follow the stellar mass distribution. 
We used the V-band data from \citet{Kormendy2009} to approximate the stellar mass distribution, which is to be corrected for the detection incompleteness to facilitate a direct comparison with the observed distribution of the field-LMXBs. 
To calculate the detection incompleteness of field-LMXBs at a given annulus, 
we assumed the broken power-law luminosity function from \citet{Kim2009}, which is based on three elliptical galaxies. 
The resultant starlight distribution is shown by the red curve in the right panel of Figure 9, which extends to 26$\arcmin$ covering the Off-Field.

We normalized the starlight distribution by matching it to the radial profile of the field-LMXBs at $\sim$1\farcm3, i.e., the first green data point in the right panel of Figure 9. 
The two inner data points in blue represent sources detected in the core, which are subject to incompleteness due to the removal of spurious sources (Section 3). 
Nevertheless, these two points appear to be consistent with the starlight distribution, indicating that the filtering of the spurious sources was reasonble. 
There is a significant excess of $\sim$30 sources between $\sim$2\farcm5--4$\arcmin$. 
We note that this radial range is where both the ACSVCS and NGVCS GCs are largely incomplete, due to limited field-of-view of the former and limited sensitivity of the latter. 
Hence, the excess seen at $\sim$2\farcm5--4$\arcmin$ could be at least partially attributed to unidentified GC-LMXBs. 
For the Off-Field, we considered only sources detected within a region that is of highest sensitivities (blue sector in Figure 1). The total number of sources detected in this region, after subtracting the CXB contribution and the 3 GC-LMXBs (Section 4.1), is $5.2(<13.6)$, 
which is consistent with the number of field-LMXBs, $\sim$3, as predicted by the stellar mass distribution.

\subsection{Variability}
\citet{Foster2013} studied eight variable X-ray point sources in the core of M87 using 85 ACIS-S observations, which span from weeks to years. However, these observations have a rather limited field-of-view ($< 10{\rm~arcmin}^{2}$). Here we search for variable sources in the larger ACIS-I field-of-view. 
We defined variable sources according to the following variation index,
\begin{eqnarray}
VI \equiv {S_{\rm max}-S_{\rm min}\over\sqrt{\sigma_{\rm max}^2+\sigma_{\rm min}^2}} \geq 3,
\end{eqnarray}
where $S_{\rm max}$ and $S_{\rm min}$ are the maximum and minimum fluxes measured among all observations that covered the given source, and $\sigma_{\rm max}$ and $\sigma_{\rm min}$ the respective errors. 
The variability thus probed is on timescales ranging from $\sim$1 day to a decade (Table 1). 
We thus identified 40 significantly variable sources, 12 of which are GC-LMXBs. The other 28 sources might be field-LMXBs or background AGNs. We provide the value of $S_{\rm max}/S_{\rm min}$ for these sources in Table 2. 6 point sources were found by \citet{Foster2013}.
Only one source (ID 274 in Table 2) exhibits $S_{\rm max}/S_{\rm min} > 10$. This source, positionally coincident with a GC, has a maximum luminosity of $\sim1.7\times10^{39}{\rm~erg~s^{-1}}$ as measured in ObsID 5826, which significantly exceeds the Eddington Luminosity of a neutron star. 
Hence, this source is likely a black hole binary in a GC, similar to the case of a GC-X-ray source in NGC\,4472 (\citealp{Maccarone2007}). 



	
\subsection{Stacking non-detected GCs and UCDs}
There must be some undetected LMXBs due to the limited sensitivity, especially in GCs. 
Hence we performed an X-ray stacking analysis for those GCs without an X-ray counterpart. 
In order to minimize contamination by the diffuse hot gas, we only use the hard band (2--8 keV), and the GCs in the core of M87 were abandoned, where there is obvious gas emission in 13 $\rm{arcmin^2}$. 864 GCs were used to stack where 338 GCs are red and 526 GCs are blue. 
To produce aggregated hard-band flux measurements, we used our custom IDL-based software package, STACKFAST (e.g., \citealp{Hickox2007}; \citealp{Chen2013}; \citealp{Goulding2017}), which is designed to systematically and efficiently combine X-ray data at known positions from multiple ACIS observations. 

We extracted 2--8~keV photons in $50\arcsec\times50\arcsec$ square regions centered on each of the GCs, along with their associated exposure maps. Specific subsets of the GCs lists have their X-ray photons coadded within circular regions characterized by the 90\% EER ($r_{90}$) of each individual ObsID. 
During stacking analyses, it is particularly important to calculate the source photon apertures on a per ObsID basis, as the ACIS PSF size varies dramatically across the field-of-view of the detector, which for differing roll angles and pointing positions can therefore result in variable 90\% EER for individual sources observed in multiple ObsIDs. 
Local background measurements were established by rescaling the remaining photon counts in the $50\arcsec\times50\arcsec$ region that are at radii greater than $1.3\times r_{90}$ (after removal of spurious or unrelated detected sources), to the area encapsulated within the $r_{90}$ aperture. 
The individual source, background and exposure measurements are then coadded, ultimately providing average background-subtracted X-ray fluxes for the stacked GC lists. 

We selected color and magnitude as the properties on which to sort the GCs and stack their X-ray emission. 
The photon flux per source for red GCs is 8.5$\pm 0.8\times10^{-8}{\rm~photon~cm^{-2}~s^{-1}~GC^{-1}}$ and it is 5.0$\pm 0.5\times10^{-8}{\rm~photon~cm^{-2}~s^{-1}~GC^{-1}}$ for blue GCs. The average flux for red GCs is almost double the flux for blue GCs. This is consistent with the result from GC hosting a detected LMXB. Converting these photon fluxes to unabsorbed  0.5--8.0 keV luminosities, we find that the average luminosity for red (blue) GCs is 2.6$\pm 0.2\times10^{37}{\rm~erg~s^{-1}~GC^{-1}}$ (1.5$\pm 0.2\times10^{37}{\rm~erg~s^{-1}~GC^{-1}}$).
The photon flux per source for bright GCs (defined as having $z$-band magnitude brighter than 21 mag) is 1.0$\pm 0.1\times10^{-7}{\rm~photon~cm^{-2}~s^{-1}~GC^{-1}}$ and it is 4.4$\pm 0.5\times10^{-8}{\rm~photon~cm^{-2}~s^{-1}~GC^{-1}}$ for faint GCs, which means that the average luminosity for bright (faint) GCs is 3.0$\pm 0.3\times10^{37}{\rm~erg~s^{-1}~GC^{-1}}$ (1.3$\pm 0.1\times10^{37}{\rm~erg~s^{-1}~GC^{-1}}$).
Expressing the GC luminosity in units of the $g$-band Solar luminosity and adopting the Solar $g$-band absolute magnitude of 5.12 (Sparke \& Gallagher 2000), 
we also derived 6.0$\pm0.5\times10^{31}{\rm~erg~s^{-1}~L_{g,\odot}}$ and  
2.1$\pm0.2\times10^{31}{\rm~erg~s^{-1}~L_{g,\odot}}$ for the red and blue GCs, respectively. 

In Section~4.1, we found no X-ray counterpart for any of the 76 UCDs within the center-field and the off-field. 
\citet{Hou2016} studied the X-ray emission from a large sample of UCDs (but not including any of the UCDs around M87) and showed that 3.3$\pm$0.8\% host an X-ray source above $5\times10^{36}{\rm~erg~s^{-1}}$, which they suggested to be most likely LMXBs. 
The absence of an X-ray counterpart in the 76 UCDs considered here seems at odds with the finding of \citet{Hou2016}, which would have predicted 2--3 X-ray counterparts, although this could be explained by our limiting sensitivity of $\sim$ $10^{37}{\rm~erg~s^{-1}}$. 
We have performed a similar stacking analysis for the 76 UCDs and obtained an average 0.5--8 keV luminosity of $1.4\pm0.1\times10^{36}{\rm~erg~s^{-1}}$ per UCD, which is comparable to the equivalent luminosity of non-detected UCDs found by \citet{Hou2016}.

\section{Summary and Discussion}
Based on the deep {\it Chandra} obseravtions studied in this work, 
we have detected 346 point-like X-ray sources in the inner $\sim$500$\arcsec$ ($\sim$40 kpc) of M87 (Table 2). 
Among these sources, we found 122 to be positionally coincident with an optically-identified GC. The remaining sources presumably comprise of field-LMXBs and background AGNs, which contribute approximately equally. 
This is one of the largest samples of GC-hosting X-ray sources, presumably LMXBs, in an external galaxy, nearly doubling the previous sample studied by \citet{Jordan2004}.   
On the other hand, we find no significant X-ray sources at the positions of 76 UCDs within the {\it Chandra} field-of-view (Section 4.1). 

We have found that 4.9$\pm$0.5\% of the HST/ACSVCS GCs and 5.9$\pm$0.9\% of the CFHT/NGVCS GCs host an X-ray source. 
This percentage becomes 7.9\%$\pm$1.2\% when only GCs brighter than m$_g$ = 22 mag were considered, where both GC catalogs are expected to be complete (Figure 3). 
This supports the view that more massive GCs have a larger probability to host an LMXB. Our stacking analysis of the individually-undetected GCs, which shows that brighter GCs have a higher average X-ray luminosity (Section 4.6), is also consistent with such a picture. 
On the other hand, red, metal-rich GCs have a $\sim$2.2 times higher fraction to host an X-ray counterpart than blue, metal-poor GCs, which is broadly consistent with that found in ETGs in the literature (Fabbiano 2006). 

After subtracting the CXB contribution, $\sim$50\% of the remaining X-ray sources are found to be associated with a GC (Section 4.1). 
This percentage is consistent within errors with the value of ($62\pm15$)\% reported by Jord\'{a}n et al.~(2004), which was based on shallower X-ray data and considered only the ACSVCS GCs. 
Therefore, we maintain the view that M87 exhibits a higher fraction of LMXBs residing in GCs than other ETGs (e.g., \citealp{Hou2016, Kim2006}), except NGC\,1399, which is also a cD galaxy hosting a large number of GCs. In this galaxy, $65\%\pm$5\% of the currently known GCs have an X-ray counterpart down to a limiting luminosity of a few $10^{37}{\rm~erg~s^{-1}}$ (\citealp{Paolillo2011}; \citealp{D'Ago2014}).  

The GC-LMXBs and field-LMXBs in M87 differ in their luminosity functions (Section 4.3). 
This disfavors the scenario in which the field-LMXBs were originated in GCs and later released into the field.
In particular, more luminous sources ($L_X \gtrsim 3\times10^{38}{\rm~erg~s^{-1}}$ tend to be found among the GC-LMXBs. 
The lack of such luminous sources in the field would require that either they have only recently been formed in GCs, or they are superposition of multiple, less luminous sources, which, once released into the field, become distinguished from each other.
However, at least one such source, ID 274, is strongly variable and most likely being a single source (Section 4.5).

The GC-LMXBs and field-LMXBs in M87 also differ in their radial distributions luminosity (Section 4.4). 
\citet{Gilfanov2004} found that the radial distribution of LMXBs in ETGs is generally consist with the stellar mass distribution, although he did distinguish field-LMXBs and GC-LMXBs. 
\citet{Kundu2002} found that the LMXB population follows the distribution of GCs rather than the stellar mass, while \citet{Kim2006} argued that GC-LMXBs are more likely to follow the starlight, which suggests that the field-LMXBs have been formed from GCs. 
As a giant elliptical and a cD galaxy, M87 exhibits a significantly steeper radial distribution of field-LMXBs than the GC-LMXBs (Figure 9). 
This is similar to the case of NGC\,1399 (\citealp{Paolillo2011}), but disagrees with the scenario of \citet{Kim2006}. 


The deep {\it Chandra} observations toward a field at $\sim$72--120 kpc northwest of the M87 center allow us to probe X-ray sources associated with the extended stellar halo of M87.
We have detected an addition of 173 sources at a projected distance of $\sim$600$\arcsec$--2000$\arcsec$ (48--160 kpc) to the northwest (Table 3), among which 6 are positionally coincident with a GC. 
While the majority of the other sources should be background AGNs, we statistically determine that $\sim$5 sources could be associated with the halo of M87, which is consistent with the expected nubmber of field-LMXBs given the underlying stellar mass (Section 4.4). 
This adds to the growing evidence of X-ray sources found around ETGs, in both isolated and group/cluster environments (\citealp{Li2010}; \citealp{Zhang2013}; \citealp{van Haaften2018}; \citealp{Hou2017}).
In particular, \citet{Hou2017} found a significant excess of X-ray sources outside the main stellar content of 80 ETGs in the Virgo cluster. They argued that at least some of these excess sources can be attributed to the intra-cluster stellar populations, i.e., the so-called diffuse intra-cluster light (ICL; \citealp{Mihos2017}). 
The X-ray sources found in the remote halo of M87 might be closely related to the intra-cluster populations.


\acknowledgements
This work is supported by the National Science Foundation of China under grants J1210039 and 11133001, and by NASA grant GO3-14131X.
The STACKFAST code was developed with support from Chandra archive grant SP8-9001A.
L.L. is grateful to the warm hospitality of Matthew Ashby, Jonathan McDowell and Qizhou Zhang during his visit to the Harvard-Smithsonian Center for Astrophysics, 
and acknowledges support from Top-notch Academic Programs Project of Jiangsu Higher Education Institutions. 
Z.L. acknowledges support from the Recruitment Program of Global Youth Experts.

\appendix 
For reference, we provide the positions of the visually identified spurious sources (Section 3) in Table~A1. 

\setcounter{table}{0}
\renewcommand\thetable{A\arabic{table}}
\begin{table}[h!]
\centering
\linespread{1.2}
{\small

\smallskip
\caption{Spurious sources}

\begin{tabular}{ccc}
\hline\noalign{\smallskip}
\hline\noalign{\smallskip}
\\

 ID& RA &Dec  \\
(1) & (2) & (3) \\

\hline\noalign{\smallskip}
1	&	12:30:34.42	&	12:19:26.61	\\
2	&	12:30:34.52	&	12:19:23.79	\\
3	&	12:30:35.60	&	12:20:18.33	\\
4	&	12:30:37.71	&	12:20:09.51	\\
5	&	12:30:37.73	&	12:20:57.72	\\
6	&	12:30:38.04	&	12:20:11.49	\\
7	&	12:30:39.08	&	12:20:33.30	\\
8	&	12:30:39.15	&	12:20:32.61	\\
9	&	12:30:39.15	&	12:21:10.02	\\
10	&	12:30:39.29	&	12:21:08.54	\\

\hline
\end{tabular}
\tablecomments{(1) Source ID; (2)-(3) RA and DEC of source centroid in epoch J2000. 
(Only a portion of the full table is shown here to illustrate its form and content.)}
}
\label{tab:spurious}
\end{table}


\newpage
\begin{thebibliography}{}
\bibitem[Angelini et al.(2001)]{Angelini2001} Angelini, L., Loewenstein, M., \& Mushotzky, R. F. 2001, ApJ, 557, L35
\bibitem[Bonnarel et al.(2000)]{Bonnarel2000} Bonnarel, F., Fernique, P., Bienaym\'{e}, O., et al. 2000, A\&AS, 143, 33B
\bibitem[Cappellari et al.(2006)]{Cappellari2006} Cappellari, M, Bacon, R., Bureau, M., et al. 2006, MNRAS, 366, 1126C
\bibitem[Chen et al.(2013)]{Chen2013} Chen, C. J., Hickox, R. C., Alberts, S., et al. 2013, ApJ, 773, 3C 
\bibitem[Cohen et al.(1998)]{Cohen1998} Cohen, J.G., Blakeslee, J.P., \& Ryzhov, A. 1998, ApJ, 496, 808
\bibitem[C\^{o}t\'{e} et al.(2001)]{Cote2001} C\^{o}t\'{e}, P., McLaughlin, D. E., Hanes, D. A., et al. 2001, ApJ, 559, 828
\bibitem[D'Ago et al.(2014)]{D'Ago2014} D'Ago, G., Paolillo, M., Fabbiano, G., et al. 2014, A\&A, 567, 2
\bibitem[Fabbiano (2006)]{Fabbiano2006} Fabbiano, G. 2006, ARA\&A, 44, 323
\bibitem[Ferrarese et al.(2012)]{Ferrarese2012} Ferrarese, L., C\^{o}t\'{e}, P., Cuillandre, J.-C., et al. 2012 ApJS, 200, 4
\bibitem[Foster et al.(2013)]{Foster2013} Foster, D. L., Charles, P. A., Swartz, D. A., et al. 2013, MNRAS, 432, 1375F
\bibitem[Georgakakis et al.(2008)]{Georgakakis2008} Georgakakis, A., Nandra, K., Laird, E. S., et al. 2008, MNRAS, 388, 1205
\bibitem[Gilfanov et al.(2004)]{Gilfanov2004} Gilfanov, M. 2004, MNRAS, 349, 146G 
\bibitem[Goulding et al.(2017)]{Goulding2017} Goulding, A. D., Matthaey, E., Greene, J. E., et al. 2017, ApJ, 843, 135G
\bibitem[Hanes et al.(2001)]{Hanes2001} Hanes, D. A., C\^{o}t\'{e}, P., Bridges, T. J., et al. 2001, ApJ, 559, 812
\bibitem[Harris et al.(2003)]{Harris2003} Harris, D. E., Biretta, J. A., Junor, W., et al. 2003, ApJ, 586L, 41H
\bibitem[Harris et al.(1991)]{Harris1991} Harris, W. E., Allwright, J. W. B., Pritchet, C., J., et al. 1991, ApJS, 76, 115H 
\bibitem[Harris et al.(1998)]{Harris1998} Harris, W. E., Harris, G.L.H., \& McLaughlin, D.E. 1998, AJ, 115, 1801
\bibitem[Hickox et al.(2007)]{Hickox2007} Hickox, R C., \& Markevitch, M., 2007, ApJ, 671, 1523H
\bibitem[Hong et al.(2009)]{Hong2009} Hong, J. S., van den Berg, M., Grindlay, J. E., 2009, ApJ, 706, 223H
\bibitem[Hou \& Li (2016)]{Hou2016} Hou, M.\& Li, Z., 2016, ApJ, 819, 164H
\bibitem[Hou et al.(2017)]{Hou2017} Hou, M.; Li, Z.; Peng, E. W., et al. 2017, ApJ, 846, 126H
\bibitem[Jarrett et al.(2003)]{Jarrett2003} Jarrett, T. H., Chester, T., Cutri, R., et al. 2003, AJ, 125, 525
\bibitem[Jord\'{a}n et al.(2003)]{Jordan2002} Jord\'{a}n, A., C\^{o}t\'{e}, P., West, M.J., et al. 2002, ApJ, 576, L113
\bibitem[Jord\'{a}n et al.(2004)]{Jordan2004} Jord\'{a}n, A., C\^{o}t\'{e}, P.; Ferrarese, L., et al. 2004, ApJ, 613, 279J
\bibitem[Jord\'{a}n et al.(2009)]{Jordan2009} Jord\'{a}n, A., Peng, E. W., Blakeslee, J. P., et al. 2009, ApJS, 180, 54J
\bibitem[Jord\'{a}n et al.(2007)]{Jordan2007} Jord\'{a}n, A. Sivakoff, G. R., McLaughlin, D. E., et al. 2007, ApJ, 671, L117
\bibitem[Kim et al.(2006)]{Kim2006} Kim, E., Kim, D.-W., Fabbiano, G., et al. 2006, ApJ, 647, 276K
\bibitem[Kim et al.(2007)]{Kim2007} Kim, M., Kim, D,-W., Wilkes, B. J., et al. 2007, ApJS, 169, 401
\bibitem[Kim et al.(2009)]{Kim2009} Kim, D.-W., Fabbiano, G., Brassington, N. J., et al. 2009, ApJ, 703, 829K
\bibitem[Kim et al.(2012)]{Kim2012} Kim, D.-W., Fabbiano, G., Pipino, A., et al. 2012, ApJ, 751, 38K
\bibitem[Kissler-Patig et al.(2002)]{Kissler-Patig2002} Kissler-Patig, M., Brodie, J.P., \& Minniti, D. 2002, A\&A, 391, 441
\bibitem[Ko et al.(2017)]{Ko2017} Ko, Y., Hwang, H., Lee, M., Park, H. S., et al. 2017, ApJ, 835, 212K
\bibitem[Kormendy et al.(2009)]{Kormendy2009} Kormendy, J., Fisher, D. B., Cornell, M. E. et al. 2009, ApJS, 182, 216K
\bibitem[Kundu et al.(2002)]{Kundu2002} Kundu, A., Majewski, S. R., Rhee, J., et al. 2002, ApJ, 576L, 125K
\bibitem[Kundu et al.(1999)]{Kundu1999} Kundu, A., Whitmore, B. C., Sparks, W. B., et al., 1999,  ApJ, 513, 733
\bibitem[Lahav et al.(1992)]{Lahav1992} Lahav, O., \& Saslaw, W. C., 1992, ApJ, 396, 430
\bibitem[Lehmer et al.(2012)]{Lehmer2012} Lehmer, B. D., Xue, Y. Q., Brandt, W. N., et al. 2012, ApJ, 752, 46L
\bibitem[Li et al.(2010)]{Li2010} Li, Z., Spitler, L. R., Jones, C., et al. 2010, ApJ, 730, 84L
\bibitem[Maccarone et al.(2007)]{Maccarone2007} Maccarone, T. J., Kundu, A., Zepf, S. E., Rhode, K. L., 2007, Nature, 445, 183
\bibitem[McLaughlin et al.(1994)]{McLaughlin1994} McLaughlin, D. E., Jarris, W. E., \& Hanes, D. A. 1994, ApJ, 422, 486
\bibitem[Mihos et al.(2017)]{Mihos2017} Mihos, J. C., Harding, P., Feldmeier, J. J., et al. 2017, ApJ, 834, 16M
\bibitem[Muno et al.(2009)]{Muno2009} Muno, M. P., Bauer, F. E., Baganoff, F. K., et al. 2009, ApJS, 181, 110
\bibitem[Park et al.(2006)]{Park2006} Park, T., Kashyap, V.L., Siemiginowska, A., et al. 2006, ApJ, 652, 610 
\bibitem[Paolillo et al.(2011)]{Paolillo2011} Paolillo, M., Puzia, T. H., Goudfrooij, P., et al. 2011, ApJ, 736, 90P
\bibitem[Powalka et al.(2016)]{Powalka2016} Powalka, M., Lan\c{c}on, A., Puzia, T. H., et al. 2016, ApJS, 227, 12P
\bibitem[Sparke et al.(2000)]{Sparke2000} Sparke, L. S., \& Gallagher, J. S. 2000, Galaxies in the Universe: An Introduction (Cambridge: Cambridge Univ. Press)
\bibitem[Sarazin et al.(2000)]{Sarazin2000} Sarazin, C. L., Erwin,J. A.,\& Bregman, J. N. 2000, ApJ, 544, L101
\bibitem[Tonry et al.(2001)]{Tonry2001} Tonry, J. L., Dressler, A., Blakeslee, J. P., 2001, ApJ, 546, 681T
\bibitem[van Haaften et al.(2018)]{van Haaften2018} van Haaften, L. M., Maccarone, T. J., Sell, P. H., et al., 2018, ApJ, 853, 13V 
\bibitem[Voss \& Gilfanov (2006)]{Voss2006} Voss, R.\& Gilfanov, M. 2006, A\&A, 447, 71V
\bibitem[Voss et al.(2009)]{Voss2009} Voss, R., Gilfanov, M., Sivakoff, G. R., et al. 2009, ApJ, 701, 471 
\bibitem[Zhang et al.(2012)]{Zhang2012} Zhang, Z., Gilfanov, M., Bogd\'{a}n, \'{A}., 2012, A\& A, 546A, 36Z
\bibitem[Zhang et al.(2013)]{Zhang2013} Zhang, Z., Gilfanov, M., Bogd\'{a}n, \'{A}., 2013, A\& A, 556A, 9Z
\bibitem[Zhu et al.(2018)]{Zhu2018} Zhu, Z., Li, Z., Morris, M. R. 2018, ApJS, 235, 26
\end{thebibliography}
\end{document}